\documentclass{cpbtex}
\usepackage{graphicx}

\usepackage{amsmath}
\usepackage{amssymb}   

\usepackage{bm}

\begin{document}

\title{Explicit forms of zero modes in symmetric interacting Kitaev chain without and with dimerization
\thanks{Project supported by the National Natural Science Foundation of China (Grant No.~11274379) and the Research Funds of Renmin University of China (Grant No. 14XNLQ07).}
}

\author{Yiming Wang$^{1}$, \ Zhidan Li$^{1}$, \ and \ Qiang Han$^{1,2}$\thanks{Corresponding author. E-mail:~hanqiang@ruc.edu.cn}\\
$^{1}${Department of Physics, Renmin University of China, Beijing 100872, China} \\
$^{2}${Beijing Key Laboratory of Opto-electronic Functional Materials and Micro-nano Devices,}\\ {Renmin University of China, Beijing 100872, China}
}

\date{\today}
\maketitle

\begin{abstract}
The fermionic and bosonic zero modes of the 1D interacting Kitaev chain at the symmetric point are unveiled. The many-body structures of the Majorana zero modes in the topological region are given explicitly by carrying out perturbation expansion up to infinite order. We also give the analytic expressions of the bosonic zero modes in the topologically trivial phase.
Our results are generalized to the hybrid fermion system comprised of the interacting Kitaev model and the Su-Schrieffer-Heeger model, in which we show that these two types of zero modes can coexist in certain region of its phase diagram.  
\end{abstract}



\maketitle

The Kitaev chain has stimulated intense research interest in the community of condensed matter physics since it was first proposed in the pioneer work of Kitaev.~\cite{Kitaev2001}  As a model of one-dimensional (1D) topological $p$-wave superconductor,  the Kitaev chain hosts two unpaired Majorana zero modes (MZMs) nonlocally distributed at the two ends of the chain.~\cite{Kitaev2001,Alicea2012,Fendley2012,Hegde2016} These MZMs are exotic quasiparticles which are their own antiparticles and topological superconductors possessing well-separated MZMs are potential platforms to realize fault tolerant topological quantum computation due to their non-Abelian statistics and immunity to local perturbations.\cite{Sau2011,Alicea2011,Leijnse2012} Signatures of observing the MZMs have been reported by several experimental groups~\cite{Mourik2012,Das2012,Deng2012,Rokhinson2012} based on theoretical proposals of realizing effective $p$-wave pairing in the spin-orbit coupled semiconductor nanowires in proximity to $s$-wave superconductors.~\cite{Fu2008,Oreg2010,Lutchyn2010,Stanescu2011}

To gain deeper insight into the MZMs beyond the single-particle picture, the interacting Kitaev chain has been studied theoretically.~\cite{Hassler2012,Thomale2012,Katsura2015,Kells2015_1,Kells2015_2,Yizhou2017}  The variation of the zero-energy peak in the local density of states was examined numerically~\cite{Thomale2012} as a reflection of the effect of interaction on the MZMs.
Futhermore, the interacting Kitaev chain at the symmetric point was shown to be exactly solvable and exact solutions of all the eigenstates were given.~\cite{Yizhou2017}
By investigating the two degenerate ground states, it was pointed out that there are fermionic or bosonic zero modes in the topologically nontrivial or trivial phase.~\cite{Yizhou2017} However, the exact structures of the zero modes have not been given explicitly for the interacting Kitaev chain even at the symmetric point, although the MZMs in the presence of interaction are expected to be adiabatically connected to the noninteracting ones.~\cite{Kitaev2001,Katsura2015}

In this paper we will present the analytic expressions of the MZMs as well as the bosonic zero modes (BZMs) in the interacting (dimerized) Kitaev chain at the symmetric point.
The main purpose of this paper is to give the many-body generalization of MZM. With the help of the explicit forms of the zero modes in the interacting (dimerized) Kitaev chain, the topological phase diagram are also given.

Before giving the exact results, we first discuss some general restrictions on the boundary zero modes in the fermion system.~\cite{Fendley2012,Katsura2015,Kells2015_1} The many-body zero mode $\hat{O}$ of the interacting Kitaev chain is a Hermitian operator which commutes with the system Hamiltonian,~\cite{Goldstein}
\begin{equation}
    [ \hat{H}, \hat{O}]  = 0, \ \ \  \hat{O}^\dagger = \hat{O}. \label{cond1}
\end{equation}
The MZM (BZM) is of the fermionic (bosonic) type and therefore anticommutes (commutes) with the parity operator of the system,
\begin{equation}
    \{ (-)^{\hat{N}}, \hat{O}_\text{F} \} = 0, \ \ \ [ (-)^{\hat{N}}, \hat{O}_\text{B} ] = 0, \label{cond2}
\end{equation}
where $\hat{N}$ is the operator of total fermion number. As a physical zero mode, $\hat{O}$ acts on a normalizable state $|\psi\rangle$ to get another normalizable state $\hat{O}|\psi\rangle$, which demands that $\hat{O}$ is unitary and thus satisfies the normalization condition,
\begin{equation} \label{normcond}
    \hat{O}_{\text{F}(B)}^2 = 1.
\end{equation}
In addition, as a boundary mode, the components of the MZM or BZM is exponentially small with the distance from the boundary.

Next we address the zero modes in the interacting Kitaev chain with open boundary condition, of which the Hamiltonian is written as,
\begin{equation} \label{hamKitaevU}
    \hat{H} = \hat{H}_\text{K}+\hat{H}_\text{I}.
\end{equation}
The system Hamiltonian is composed of a single-particle part $\hat{H}_\text{K}$ and a fermion-fermion interaction part $\hat{H}_\text{I}$.
$\hat{H}_\text{K}$ denotes the tight-binding Hamiltonian of 1D Kitaev chain, which is written as
\begin{equation} \label{HK}
    \hat{H}_\text{K} = -\sum_{n=1}^{N-1} ( t c_n^\dagger c_{n+1} + \Delta c_n^\dagger c_{n+1}^\dagger + h.c. ) - \mu \sum_{n} c_n^\dagger c_{n},
\end{equation}
with $t$ the nearest-neighbor hopping integral, $\Delta$ the $p$-wave superconducting pairing potential, and $\mu$ the chemical potential. $N$ denotes the length of the chain. $c_n$ and $c_n^\dagger$ are the fermion annihilation and creation  operators, respectively. $H_\text{I}$ is expressed as
\begin{equation} \label{HI}
    \hat{H}_\text{I}= U \sum_{n=1}^{N-1} \left(c_n^\dagger c_n -\frac{1}{2} \right) \left(c_{n+1}^\dagger c_{n+1} - \frac{1}{2} \right),
\end{equation}
where $U$ is the interaction between nearest-neighbor fermions.

In the following, we focus on the symmetric case where $t=\Delta$ and $\mu=0$, and the thermodynamic limit $N\to\infty$ is taken. Without loss of generality we set $t>0$. For later convenience, we rewrite Eq.~(\ref{hamKitaevU}) in the Majorana basis and into a dimensionless form,
\begin{equation} \label{ham}
    \hat{h} = \hat{H}/t=\hat{h}_0 + u \hat{h}_1,
\end{equation}
where $u= {U}/{4t}$ is a dimensionless parameter describing the relative strength of the interaction.
\begin{align}
    & \hat{h}_0 = i\sum_{n=1}^{N-1} b_{n} a_{n+1}, \\
    & \hat{h}_1 = -\sum_{n=1}^{N-1} a_{n} b_{n} a_{n+1} b_{n+1}.
\end{align}
Here $a_{n}=c_n+c_n^\dagger$ and $b_{n}=-i(c_n-c_n^\dagger)$ are Majorana fermion operators which satisfy $a_{n}^\dagger=a_{n}$, $b_{n}^\dagger=b_{n}$, and the anti-commutation relations $\{a_m, a_n\}=2\delta_{m,n}$, $\{b_m, b_n\}=2\delta_{m,n}$ and $\{a_m,b_n\}=0$.
When $u=0$, it was shown by Kitaev~\cite{Kitaev2001} that there is an exact MZM at each end of the chain, namely $[\hat{h}_0, a_1]=[\hat{h}_0, b_{N}]=0$, indicating that
the MZMs are composed of one-body Majorana fermions. In the presence of interaction, many-body contributions are involved and higher order terms occur in the expression of $\hat{O}_\text{F}$. For small $u$, we can treat $u\hat{h}_1$ in Eq.~(\ref{ham}) as a small quantity and expand $\hat{O}_\text{F}$ into power series of $u$ as follows,
\begin{equation} \label{OFexpansion}
    \hat{O}_\text{F}= C_\text{F}\sum_{n=0} u^n \eta_n,
\end{equation}
where $C_\text{F}$ is the normalization factor and the coefficients $\eta_n$'s are operators to be determined. Substituting Eq.~(\ref{OFexpansion}) into $[\hat{h},\hat{O}_\text{F}]=0$ and comparing the coefficient of $u^n$, we find that
$\eta_n$ satisfies the following recurrence relation,
\begin{equation}
    [\hat{h}_0, \eta_n] + [\hat{h}_1, \eta_{n-1}]=0.
\end{equation}
Starting from the leading term $\eta_0=a_1$, one can obtain $\eta_n$ successively and the general expression is,
\begin{equation} \label{etan}
    \eta_n = a_{2n+1} \prod_{m=1}^{n} i b_{2m-1} a_{2m}.
\end{equation}
$\eta_n$'s are many-body Majorana fermion operators which satisfy $\eta_n^\dagger=\eta_n$, $\{\eta_n,\eta_{n'}\}=2\delta_{n,n'}$. Substituting Eq.~(\ref{etan}) into Eq.~(\ref{OFexpansion}), we have the analytic expression of the zero mode,
\begin{equation} \label{fl}
    \hat{O}_\text{F}^l = C_\text{F} \sum_{n=0} u^n a_{2n+1} \prod_{m=1}^{n} ib_{2m-1} a_{2m},
\end{equation}
which starts from the left-most Majorana fermion $a_1$ and decays exponentially with the distance $n$ on condition that $|u|<1$.
Similarly, one can obtain the right-boundary zero mode starting from $b_N$,
\begin{equation} \label{fr}
    \hat{O}_\text{F}^r = C_\text{F} \sum_{n=0} u^n b_{N-2n} \prod_{m=1}^{n} ia_{N-2m+2} b_{N-2m+1}.
\end{equation}
$\hat{O}_\text{F}^{l}$ and $\hat{O}_\text{F}^{r}$  satisfy Eqs.(\ref{cond1}), (\ref{cond2}) and the anti-commutation relation $\{\hat{O}_\text{F}^l,\hat{O}_\text{F}^r\}=0$. To guarantee $\hat{O}_\text{F}^{l(r)}$ being MZM, the normalization condition Eq.~(\ref{normcond}) must be satisfied. From Eq.~(\ref{normcond}) we obtain $u^2<1$, or equivalently $|U|<4t$. The normalization factor is $C_\text{F}=\sqrt{1-u^2}$. Therefore the many-body MZM emerges in the topological region $|U|<4t$.~\cite{Yizhou2017} In addition, according to Eqs.~(\ref{fl}) and (\ref{fr}) the many-body MZMs are adiabatically connected to the one-body MZMs $a_1$ and $b_N$ as $U\to 0$.

In the topologically trivial region $u^2>1$, on the other hand, the many-body MZMs do not exist. On the contrary we find BZMs localized at ends of the chain. Note that when $t=0$, i.e. $u\to\infty$, there are two bosonic boundary zero modes $ia_1b_1$ and $ia_Nb_N$ satisfying $[\hat{h}_1, ia_1b_1]=[\hat{h}_1, ia_Nb_N]=0$. For large but finite $u$, one can treat $\hat{h}_0$ in Eq.~(\ref{ham}) as perturbation and expand $\hat{O}_\text{B}$ as power series of $u^{-1}$,
\begin{equation} \label{OBexpansion}
    \hat{O}_\text{B}= C_\text{B} \sum_{n=0} u^{-n} \xi_n,
\end{equation}
where $C_\text{B}$ is the normalization factor. Substituting Eq.~(\ref{OBexpansion}) into $[\hat{h},\hat{O}_\text{B}]=0$ and comparing the coefficient of $u^{-n}$, we find that
$\xi_n$ satisfies the following recurrence relation,
\begin{equation}
    [\hat{h}_0, \xi_n] + [\hat{h}_1, \xi_{n+1}]=0.
\end{equation}
Starting from the first term $\xi_0=ia_1b_1$, one can obtain $\xi_n$ successively and the general expression is,
\begin{equation} \label{xin}
    \xi_n = ia_{2n+1}b_{2n+1} \prod_{m=1}^{n} i a_{2m-1} b_{2m}.
\end{equation}
Similar to $\eta_n$, $\xi_n$'s are also many-body Majorana fermion operators which satisfy $\xi_n^\dagger=\xi_n$, $\{\xi_n,\xi_{n'}\}=2\delta_{n,n'}$. Substituting Eq.~(\ref{xin}) into Eq.~(\ref{OBexpansion}), we have the expression for the left-boundary  zero mode,
\begin{equation} \label{bl}
    \hat{O}_\text{B}^l = C_\text{B} \sum_{n=0} u^{-n} i a_{2n+1}b_{2n+1} \prod_{m=1}^{n} ia_{2m-1} b_{2m}.
\end{equation}
Likewise, we obtain the right-boundary  zero mode starting from $ia_Nb_N$,
\begin{equation} \label{br}
    \hat{O}_\text{B}^r = C_\text{B} \sum_{n=0} u^{-n} i a_{N-2n}b_{N-2n} \prod_{m=1}^{n} ia_{N-2m+1} b_{N-2m+2}.
\end{equation}
Eqs.~(\ref{bl}) and (\ref{br}) indicate that $\hat{O}_\text{B}^{l}$ and $\hat{O}_\text{B}^{r}$ are BZMs and in additon they satisfy the
commutation relation $[\hat{O}_\text{B}^l,\hat{O}_\text{B}^r]=0$. Furthermore from Eq.~(\ref{normcond}), $\hat{O}_\text{B}^{l(r)}$ is normalizable under the condition $u^2>1$, which shows
that the two BZMs exist only in the topologically trivial region $|U|>4t$.  The corresponding normalization factor is $C_\text{B}=\sqrt{1-u^{-2}}$.

\begin{figure}[ht]
\begin{center}
  \includegraphics[width=6.0cm, angle=0]{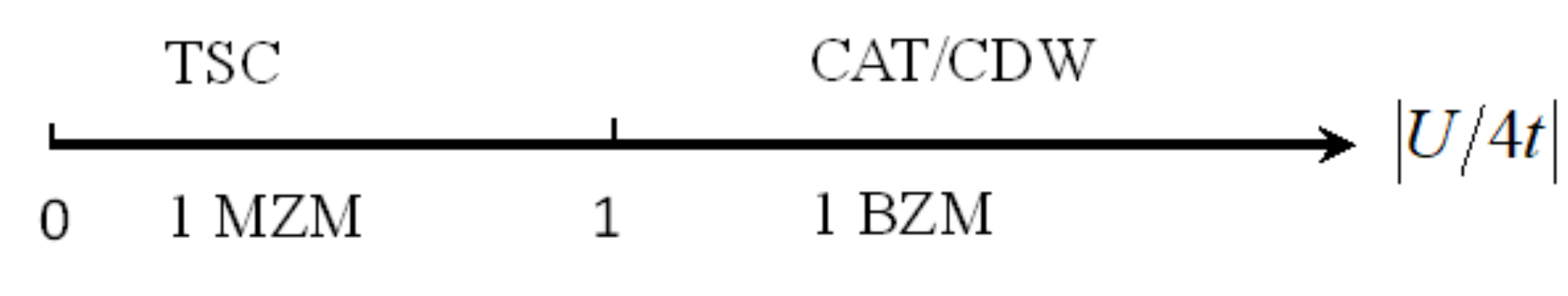}
\end{center}
 \caption{(Color online) Phase diagram of the interacting Kitaev chain at the symmetric point with $t=\Delta$ and $\mu=0$, where the topologically nontrivial (trivial) phase in the region $|U|<4t$ ($|U|>4t$) hosts one MZM (BZM)  at each end of the chain.}
 \label{fig1}
\end{figure}
The corresponding conditions for the emergence of different types of zero modes can be employed to depict the phase diagram of the symmetric interacting Kitaev chain as shown in Fig.~1 which is identical with that obtained by exact diagonalization of the system Hamiltonian.~\cite{Yizhou2017}

\begin{figure}[ht]
\begin{center}
  \includegraphics[width=7.0cm, angle=0]{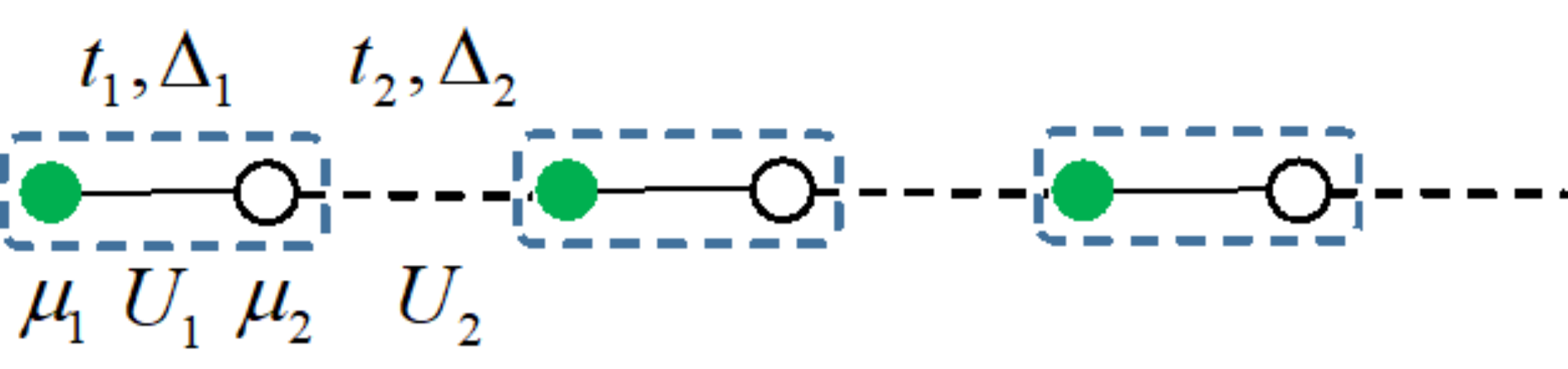}
\end{center}
 \caption{Illustration of the interacting Kitaev-SSH chain model. Dashed squares denote the unit cells and green solid (empty) circles denote sites on sublattice C (D). See the main text for more details.}
 \label{fig2}
\end{figure}
We next discuss the coexistence of both the MZMs and the BMZs in a hybrid system which comprises of the interacting Kitaev chain model and the Su-Schrieffer-Heeger (SSH) model~\cite{SSH} as shown in Fig.\ref{fig2}, which is a dimerized generalization of the interacting Kitaev chain studied above. The noninteracting version of this model has been studied extensively in the literature.~\cite{NagaosaSSH2014,Sticlet2014,Gao2015,Daping2016,zhoubozhen2016,Klett2017,Lin2017,Kawabata2017} This so-called interacting Kitaev-SSH chain consists of $N$ unit cells each of which hosts two inequivalent lattice site, $C$ and $D$.
The model Hamiltonian is written as,
\begin{equation}\label{H0}
    H = H_\text{K-SSH} + H_\text{I},
\end{equation}
where
\begin{align}
    H_\text{K-SSH} & =  -\sum_{n=1}^{N-1} ( t_1 c_n^\dag d_n +  t_2 d_n^\dag c_{n+1} + h.c. ),  \nonumber \\
                     & - \sum_{n=1}^{N-1} ( \Delta_1 c_n^\dagger d_n^\dagger + \Delta_2 d_n^\dagger c_{n+1}^\dagger + h.c. ) \nonumber \\
                     & - \sum_{n=1}^N ( \mu_1 c_n^\dagger c_n + \mu_2 d_n^\dagger d_n)
\end{align}
and
\begin{align}
    H_\text{I} & = U_1 \sum_{n=1}^{N} \left(c_n^\dagger c_n -\frac{1}{2} \right) \left(d_{n}^\dagger d_{n} - \frac{1}{2} \right), \nonumber \\
                 & + U_2 \sum_{n=1}^{N-1} \left(d_n^\dagger d_n -\frac{1}{2} \right) \left(c_{n+1}^\dagger c_{n+1} - \frac{1}{2} \right).
\end{align}
Here $n$ is the index of unit cell. $c_n$($c_n^\dagger$)  and $d_n$($d_n^\dagger$) denote the annihilation (creation) operators of fermion on the site of the sublattice $C$ ($D$), respectively. The intra(inter)-unit-cell hopping integral, pairing potential, and fermion-fermion interaction are denoted by $t_{1(2)}$, $\Delta_{1(2)}$, and $U_{1(2)}$, respectively. Hereafter, we focus on the symmetric point with $t_1=\Delta_1$, $t_2=\Delta_2$ and $\mu_1=\mu_2=0$.

\begin{figure}[ht]
\begin{center}
  \includegraphics[width=7.0cm, angle=0]{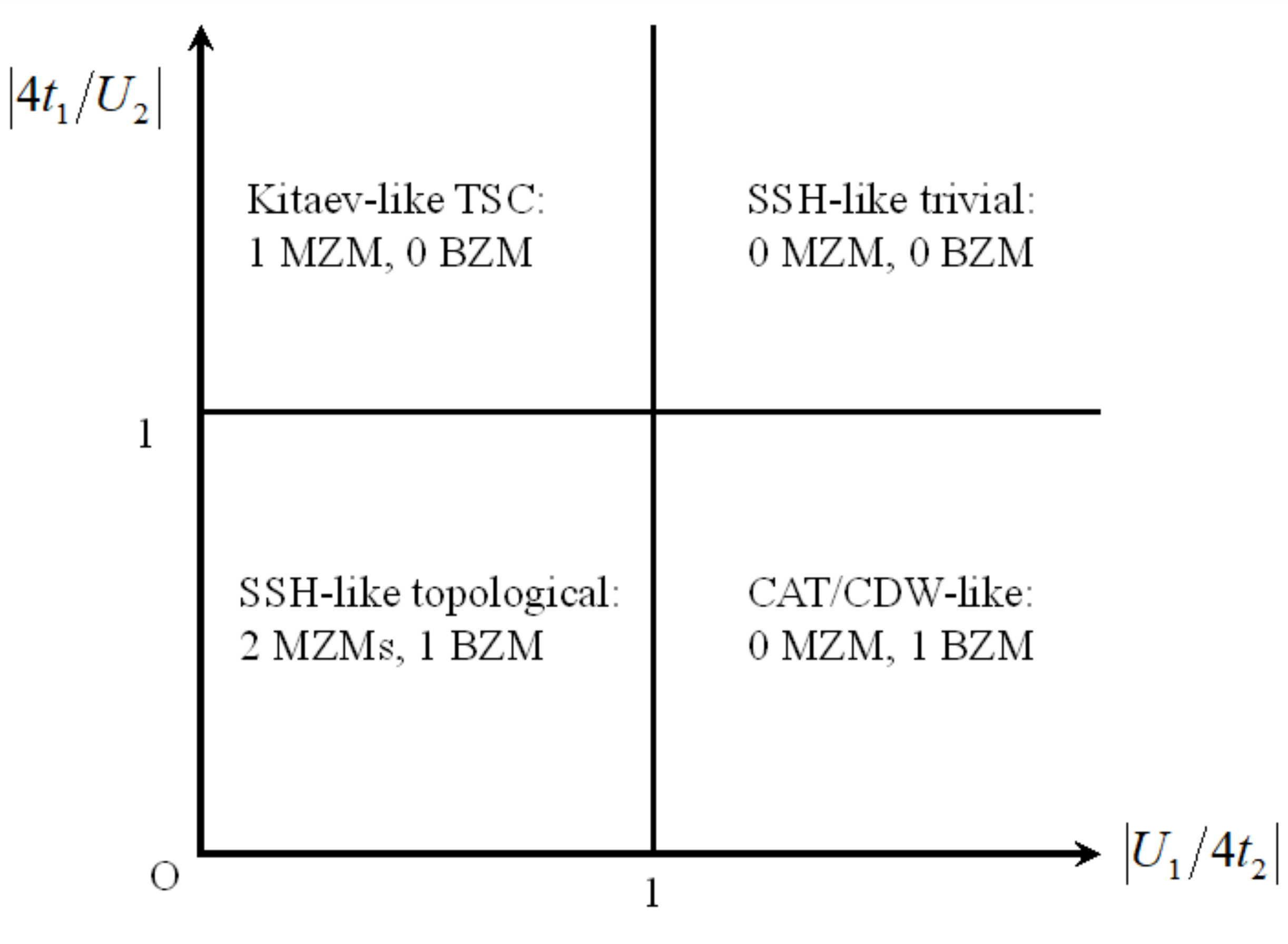}
\end{center}
 \caption{Phase diagram of the symmetric interacting Kitaev-SSH chain model with $t_1=\Delta_1$, $t_2=\Delta_2$ and $\mu_1=\mu_2=0$.}
 \label{fig3}
\end{figure}
The Majorana fermion operators are defined as $a_n=c_n+c_n^\dagger$, $b_n=-i(c_n-c_n^\dagger)$, $\alpha_n=d_n+d_n^\dagger$, $\beta_n=-i(d_n-d_n^\dagger)$.
$a_n, b_n$ are associated with the $C$ sublattice while $\alpha_n, \beta_n$ with $D$. In the Majorana representation, we have
\begin{equation}
    H_{\text{K-SSH}} = i\sum_n (t_1b_n\alpha_n + t_2\beta_na_{n+1}),
\end{equation}
and
\begin{equation}
    H_\text{I} = - \sum_n (U_1 a_n b_n\alpha_n\beta_n + U_2\alpha_n\beta_na_{n+1}b_{n+1}).
\end{equation}
Applying the forgoing method of perturbation expansion, we obtain two types of zero modes. The left-boundary MZM is written as
\begin{align} \label{flKSSH}
    \hat{O}_\text{F}^l = C_\text{F} \sum_{n=0} \left(\frac{U_1}{4t_2}\right)^n a_{n+1} \prod_{m=1}^n ib_m \alpha_m,
\end{align}
which emerges under the condition that $|U_1|<4|t_2|$, and the left-boundary BZM is
\begin{align} \label{blKSSH}
    \hat{O}_\text{B}^l = C_\text{B} \sum_{n=0} \left(\frac{4t_1}{U_2}\right)^n ia_{n+1} b_{n+1} \prod_{m=1}^n ia_m \beta_m.
\end{align}
whose existence is guaranteed by the condition that $4|t_1|<|U_2|$. The corresponding right-boundary zero modes can be obtained similarly (not shown here). One can readily check
that the when $t_1=t_2=t$, $U_1=U_2=U$, i.e. the dimerization is absent, the former equations (\ref{fl}) and (\ref{bl}) are recovered from Eqs.~(\ref{flKSSH}) and (\ref{blKSSH}), respectively. From the explicit forms of the zero modes, we derive the phase diagram of the interacting Kitaev-SSH chain as illustrated in Fig.~\ref{fig3}.
We find four phases:
(i) $|U_1|>4|t_2|$, $4|t_1|>|U_2|$, where no boundary zero modes exist; (ii) $|U_1|<4|t_2|$, $4|t_1|>|U_2|$, where there is one MZM at each end of the chain and no BZM; (iii) $|U_1|>4|t_2|$, $4|t_1|<|U_2|$, where there is one BZM at each end of the chain and no MZM; (iv) $|U_1|<4|t_2|$, $4|t_1|<|U_2|$, where
the MZM and the BZM coexist at each end of the chain. In this coexistence region, considering that $\{\hat{O}_\text{F}^l,\hat{O}_\text{B}^l\}=0$, we can construct another MZM by
\begin{equation}
    \hat{O}_\text{F}^{l\prime} = i \hat{O}_\text{F}^l \hat{O}_\text{B}^l,
\end{equation}
which fulfills Eqs.~(\ref{cond1}), (\ref{cond2}), and (\ref{normcond}) and anticommutates with $\hat{O}_\text{F}^l$. Therefore in the coexistence region, there are two MZMs at each end
of the chain as illustrated in Fig.~3. Furthermore one can construct a complex fermion by pairing these two MZMs at the same end, $\hat{C}=\hat{O}_\text{F}^l+i\hat{O}_\text{F}^{l\prime}$,
and forms a many-body SSH-like~\cite{NagaosaSSH2014} zero mode.

In summary, the many-body structures of the MZMs and BZMs in the symmetric interacting (dimerized) Kitaev chain have been investigated in this paper.  From the explicit forms of the zero modes, the many-body MZMs of the interacting model are adiabatically connected to the one-body MZMs of the noninteracting one. For the symmetric interacting Kitaev chain without dimerization, the MZMs and BZMs are found in different regions of the phase diagram. The dimerized generalization of the model has four phases depending on the numbers of zero modes and we find that both types of zero modes can coexist with each other in certain region of its phase diagram.



\begin{thebibliography}{99}


\bibitem{Kitaev2001}
Kitaev A~Y  \href{http://stacks.iop.org/1063-7869/44/i=10S/a=S29}{2001 \emph{Phys.-Usp.} \textbf{44} 131}

\bibitem{Alicea2012}
Alicea J
  \href{http://stacks.iop.org/0034-4885/75/i=7/a=076501}{2012 \emph{Rep. Prog. Phys.} \textbf{75} 076501}


\bibitem{Fendley2012}
Fendley P
  \href{http://stacks.iop.org/1742-5468/2012/i=11/a=P11020}{2012 \emph{J. Stat. Mech.}
  \textbf{2012} 11020}

\bibitem{Hegde2016}
Hegde S~S and Vishveshwara S
  \href{https://link.aps.org/doi/10.1103/PhysRevB.94.115166}{2016 \emph{Phys. Rev. B} \textbf{94} 115166}


\bibitem{Sau2011}
Sau J~D, Clarke D~J and Tewari S
  \href{https://link.aps.org/doi/10.1103/PhysRevB.84.094505}{2011 \emph{Phys. Rev. B} \textbf{84} 094505}

\bibitem{Alicea2011}
Alicea J, Oreg Y, Refael G, von Oppen F and Fisher M~P~A \href{http://dx.doi.org/10.1038/nphys1915}{2011 \emph{Nat. Phys.}
  \textbf{7} 412}

\bibitem{Leijnse2012}
Leijnse M and Flensberg K
  \href{http://stacks.iop.org/0268-1242/27/i=12/a=124003}{2012 \emph{Semicond. Sci. Tech.}
  \textbf{27} 124003}

\bibitem{Mourik2012}
Mourik V, Zuo K, Frolov S~M, Plissard S~R,
Bakkers E~P~A~M, and Kouwenhoven L~P \href{http://science.sciencemag.org/content/336/6084/1003}{2012 \emph{Science} \textbf{336}
  1003}

\bibitem{Das2012}
Das A, Ronen Y, Most Y, Oreg Y, Heiblum M, and
Shtrikman H \href{https://www.nature.com/articles/nphys2479.pdf}{2012 \emph{Nat. Phys.} \textbf{8} 887}

\bibitem{Deng2012}
Deng M~T, Yu C~L, Huang G~Y, Larsson M, Caroff P, and
Xu H~Q \href{http://pubs.acs.org/doi/ipdf/10.1021/nl303758w}{2012 \emph{Nano. Lett.} \textbf{12} 6414}

\bibitem{Rokhinson2012}
Rokhinson L~P, Liu X, and Furdyna J~K \href{https://www.nature.com/articles/nphys2429}{2012 \emph{Nat. Phys.} \textbf{8} 795}

\bibitem{Fu2008}
Fu L and Kane C~L \href{https://journals.aps.org/prl/abstract/10.1103/PhysRevLett.100.096407}{2008 \emph{Phys. Rev. Lett.} \textbf{100} 096407}

\bibitem{Oreg2010}
Oreg Y, Refael G, von Oppen F \href{https://journals.aps.org/prl/abstract/10.1103/PhysRevLett.105.177002}{2010 \emph{Phys. Rev. Lett.} \textbf{105} 177002}

\bibitem{Lutchyn2010}
Lutchyn R~M, Sau J~D and Das~Sarma S \href{https://journals.aps.org/prl/abstract/10.1103/PhysRevLett.105.077001}{2010 \emph{Phys. Rev. Lett.} \textbf{105}
  077001}

\bibitem{Stanescu2011}
Stanescu T~D, Lutchyn R~M and Das~Sarma S \href{https://link.aps.org/doi/10.1103/PhysRevB.84.144522}{2011 \emph{Phys. Rev. B} \textbf{84}
  144522}

\bibitem{Hassler2012}
Hassler F and Schuricht D
\href {https://arxiv.org/abs/1206.2134}{2012 \emph{New J. Phys.} \textbf{14} 125018}.

\bibitem{Thomale2012}
Thomale R, Rachel S, and P. Schmitteckert
\href {http://journals.aps.org/prb/pdf/10.1103/PhysRevB.88.161103}{2012 \emph{Phys. Rev. B} \textbf{88} 161103(R)}

\bibitem{Katsura2015}
Katsura H, Schuricht D and Takahashi M
\href{https://journals.aps.org/prb/pdf/10.1103/PhysRevB.92.115137}{2015 \emph{Phys. Rev. B} \textbf{92} 115137}

\bibitem{Kells2015_1}
Kells G \href{https://journals.aps.org/prb/pdf/10.1103/PhysRevB.92.155434}{2015 \emph{Phys. Rev. B} \textbf{92}  155434}

\bibitem{Kells2015_2}
Kells G  \href{https://journals.aps.org/prb/pdf/10.1103/PhysRevB.92.081401}{2015 \emph{Phys. Rev. B} \textbf{92}  081401}

\bibitem{Yizhou2017}
Miao J~J, Jin H~K, Zhang F~C and Zhou Y
    \href {https://journals.aps.org/prl/abstract/10.1103/PhysRevLett.118.267701}{2017 \emph{Phys. Rev. Lett.} \textbf{118} 267701}


\bibitem{Goldstein}
Goldstein G and Chamon Claudio \href{https://journals.aps.org/prb/pdf/10.1103/PhysRevB.86.115122}{2012 \emph{Phys. Rev. B} \textbf{86} 115122}

\bibitem{SSH}
Su W~P, Schrieffer J~R and Heeger A~J
    \href{https://journals.aps.org/prl/pdf/10.1103/PhysRevLett.42.1698}{1979 \emph{Phys. Rev. Lett.} \textbf{42} 1698}
    

\bibitem{NagaosaSSH2014}
Wakatsuki R, Ezawa M, Tanaka Y and Nagaosa N
  \href{https://journals.aps.org/prb/pdf/10.1103/PhysRevB.90.014505}{2014 \emph{Phys. Rev. B} \textbf{90} 014505}

\bibitem{Sticlet2014}
Sticlet D, Seabra L, Pollmann F and Cayssol J \href{https://journals.aps.org/prb/pdf/10.1103/PhysRevB.89.115430}{2015 \emph{Phys. Rev. B} \textbf{89}  115430}

\bibitem{Gao2015}
Gao Y, Zhou T, Huang H~X and Huang R \href{https://www.nature.com/articles/srep17049}{2015 \emph{Sci. Rep.} \textbf{5} 17049}

\bibitem{Daping2016}
Liu D~P,
  \href{http://dx.doi.org/10.1088/1674-1056/25/5/057101}{2016 \emph{Chin. Phys. B} \textbf{5} 057101}

\bibitem{zhoubozhen2016}
Zhou B Z and Zhou B \href{http://cpb.iphy.ac.cn/CN/abstract/article_68247.shtml}{2016 \emph{Chin. Phys. B} \textbf{25}  107401}

\bibitem{Klett2017}
Klett M, Cartarius H, Dast D, Main J and Wunner G  \href{https://journals.aps.org/pra/pdf/10.1103/PhysRevA.95.053626}{2017 \emph{Phys. Rev. A} \textbf{95}  053626}

\bibitem{Lin2017}
Lin Y, Hao W~C, Wang M, Qian J~Q and Guo H M \href{https://www.nature.com/articles/s41598-017-09160-x}{2017 \emph{Sci. Rep.} \textbf{7} 9210}

\bibitem{Kawabata2017}
Kawabata K, Kobayashi R, Wu N and Katsura H \href{https://journals.aps.org/prb/pdf/10.1103/PhysRevB.95.195140}{2017 \emph{Phys. Rev. B} \textbf{95}  195140}




\end{thebibliography}
\end{document}